%% file: main.tex
\documentclass[letterpaper,10pt,conference]{ieeeconf} 
\let\labelindent\relax
\IEEEoverridecommandlockouts                          
\usepackage{cite}
\usepackage{amsfonts}
\usepackage{amsmath}
\usepackage{mathrsfs}
\usepackage{algorithm}
\usepackage{algpseudocode}
\usepackage[font=small]{caption}
\usepackage{booktabs,tabularx}
\usepackage{hyperref}
\usepackage[draft]{todonotes}
\usepackage{caption}
\usepackage{subcaption}
\usepackage{graphicx}
\usepackage[normalem]{ulem}

\usepackage{eufrak}
\usepackage{bbm}
\usepackage[sans]{dsfont}
\usepackage{bm}

\usepackage{tikz}
\usepackage{pgfplots}
\usepackage{pgfplotstable,booktabs}
\usepgfplotslibrary{dateplot,fillbetween}
\usepackage{pgf}
\usepgfplotslibrary{colorbrewer}
\usepackage{lscape}

\usepackage{siunitx}
\usepackage{comment}

\usepackage{amsthm}

\theoremstyle{definition}

\DeclareMathOperator*{\argmin}{arg\,min}
\DeclareMathOperator{\Han}{\mathfrak{H}}

\IEEEoverridecommandlockouts                              
\title{\LARGE \bf
Lessons Learned from Data-Driven Building Control Experiments: \\ Contrasting Gaussian Process-based MPC, Bilevel DeePC, \\ and Deep Reinforcement Learning}

\author{Loris Di Natale*, Yingzhao Lian*, Emilio T. Maddalena*, Jicheng Shi and Colin N. Jones
\thanks{L. Di Natale, Y. Lian, E. T. Maddalena, J. Shi and C. N. Jones are with the Automatic Control Laboratory, \'Ecole Polytechnique F\'ed\'erale de Lausanne (EPFL), 1015 Lausanne, Switzerland. E-mails: {\tt\small \{loris.dinatale, yingzhao.lian, emilio.maddalena, jicheng.shi, colin.jones\}@epfl.ch}. L. Di Natale is with the Urban Energy Systems Laboratory, Swiss Federal Laboratories for Materials
Science and Technology (Empa), 8600 D\"{u}bendorf, Switzerland. This work has received support from the Swiss National Science Foundation under the RISK project (Risk Aware Data Driven Demand Response, grant number 200021 175627), and under NCCR Automation (grant
agreement 51NF40\_180545).}%
\thanks{*These authors contributed equally to this work. (Corresponding author: Yingzhao Lian)}%
}

\begin{document}

\maketitle
\thispagestyle{empty}
\pagestyle{empty}

\begin{abstract}
This manuscript offers the perspective of experimentalists on a number of modern data-driven techniques: model predictive control relying on Gaussian processes, adaptive data-driven control based on behavioral theory, and deep reinforcement learning. These techniques are compared in terms of data requirements, ease of use, computational burden, and robustness in the context of real-world applications. Our remarks and observations stem from a number of experimental investigations carried out in the field of building control in diverse environments, from lecture halls and apartment spaces to a hospital surgery center. The final goal is to support others in identifying what technique is best suited to tackle their own problems. 

\end{abstract} \smallbreak

\section{Introduction}

Many control techniques fall under the umbrella of the not strictly defined ``data-driven control'' and ``learning-based control'' categories \cite{hou2013model,benosman2018model}. A fair number of successful implementations of these methods in different domains show that such approaches are well suited to operate physical systems while attaining high performance \cite{torrente2021_gpmpc,carlet2022_deepc}. 
Most of these methods aim at reducing or eliminating the engineering burden of modelling the physical system one intends to control (in the white-box sense), leveraging data to characterize system dynamics in a more general way. 
This is in contrast with \textit{direct} data-driven control methods, which improve the control performance without any plant characterization \cite{formentin2016direct}.

Among the numerous linear modelling paradigms~\cite{ljung1998system}, behavioral theory offers a way to implicitly characterize the system dynamics based on past data~\cite{willems1997introduction}. Many nonlinear parametric models are extensions of linear ones~\cite{schoukens2019nonlinear}, such as local linear model trees~\cite{nelles2002nonlinear}, Uryson operators~\cite{gallman1975iterative}, or Hammerstein models~\cite{billings1979non}. These methods typically enforce nonlinearity using polynomial convolution \cite{gallman1971identification}

or explicit nonlinear structures~\cite{schoukens2019nonlinear}. While these standard approaches suffer from the infamous curse of dimensionality, deep Neural Network (NN) models have recently become a popular choice thanks to their scalability, flexibility, and strong representative power, and were already successfully used to characterize building temperature dynamics, e.g. in~\cite{di2021physically}. Due to their ability to quickly learn objectives and dynamics even with little data, non-parametric models have also 
drawn the attention of system and control researchers. Gaussian Processes (GPs) are arguably the most popular choice thanks to their inherent uncertainty quantification~\cite{williams2006gaussian,torrente2021_gpmpc}. 

Once a model has been chosen, it is usually used either within receding horizon predictive controllers online, based on forecasts, like in typical Model Predictive Control (MPC) schemes, or to learn explicit control laws offline, e.g. through Deep Reinforcement Learning (DRL)~\cite{sutton2018reinforcement}. Both techniques were already successfully deployed to control the temperature in real buildings, e.g. in~\cite{bunning2020experimental, yang2020model, di2021deep, svetozarevic2022data, touzani2021controlling, du2022demonstration}. Since any combination of a model and a decision maker mentioned above might be chosen, there exist many flavors of data-driven building control, as reviewed e.g. in~\cite{park2018comprehensive,maddalena2020data,wang2020reinforcement}.

\begin{figure*}[t]
\begin{subfigure}{0.528\textwidth}
    \includegraphics[width=1\textwidth]{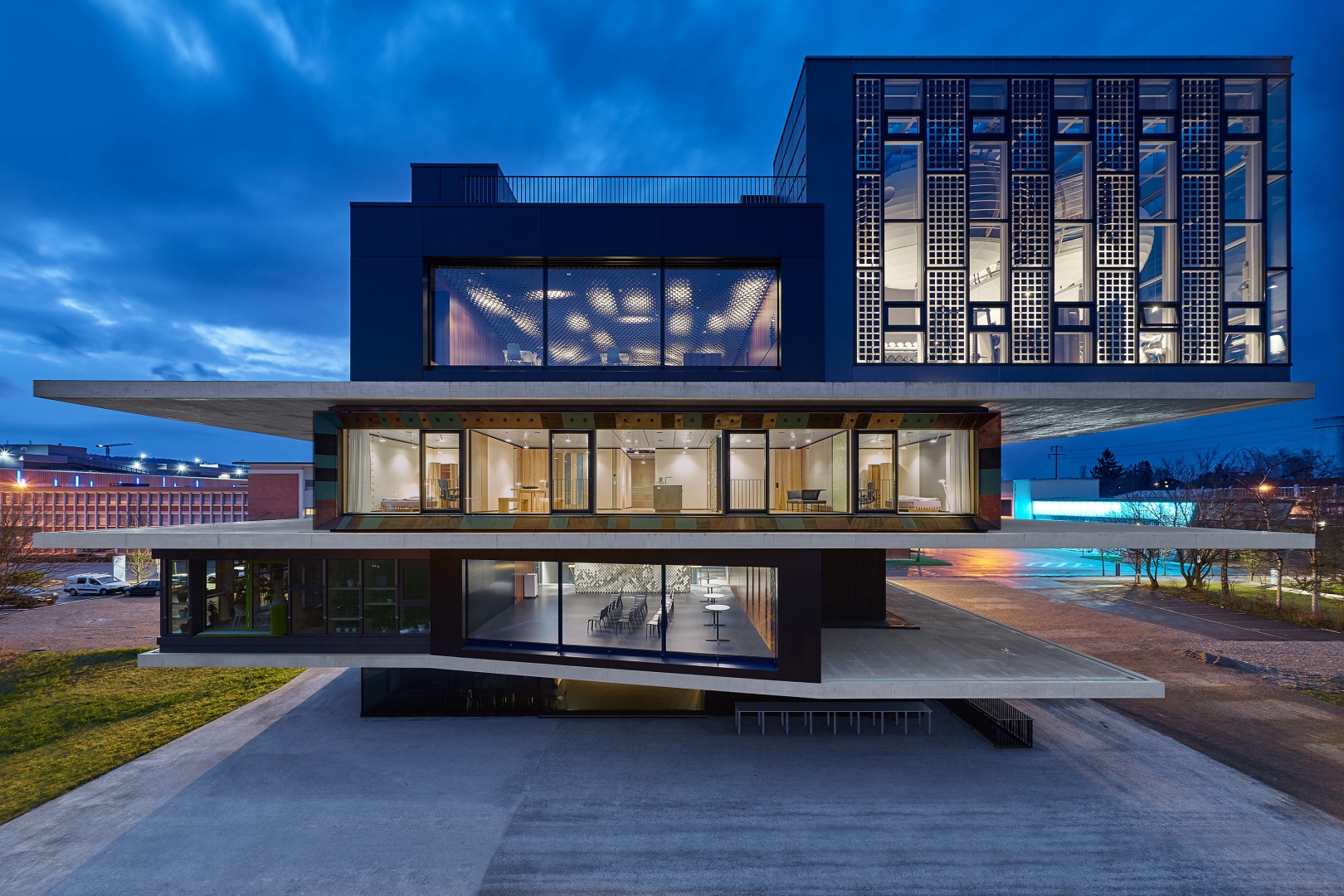}
\end{subfigure}
\hspace{0.01pt}
\begin{subfigure}[t]{0.428\textwidth}
    \begin{minipage}{1\textwidth}
    \includegraphics[width=1\textwidth]{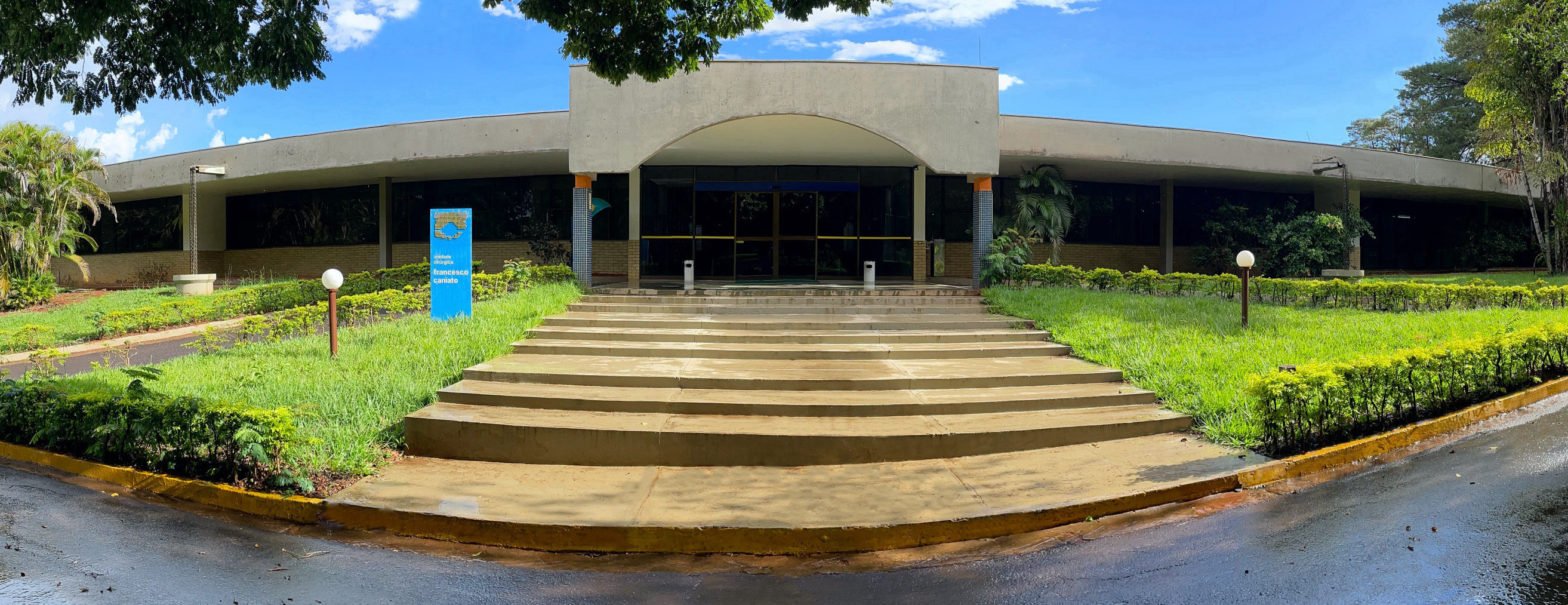}\\[6.5pt]
    \includegraphics[width=1\textwidth]{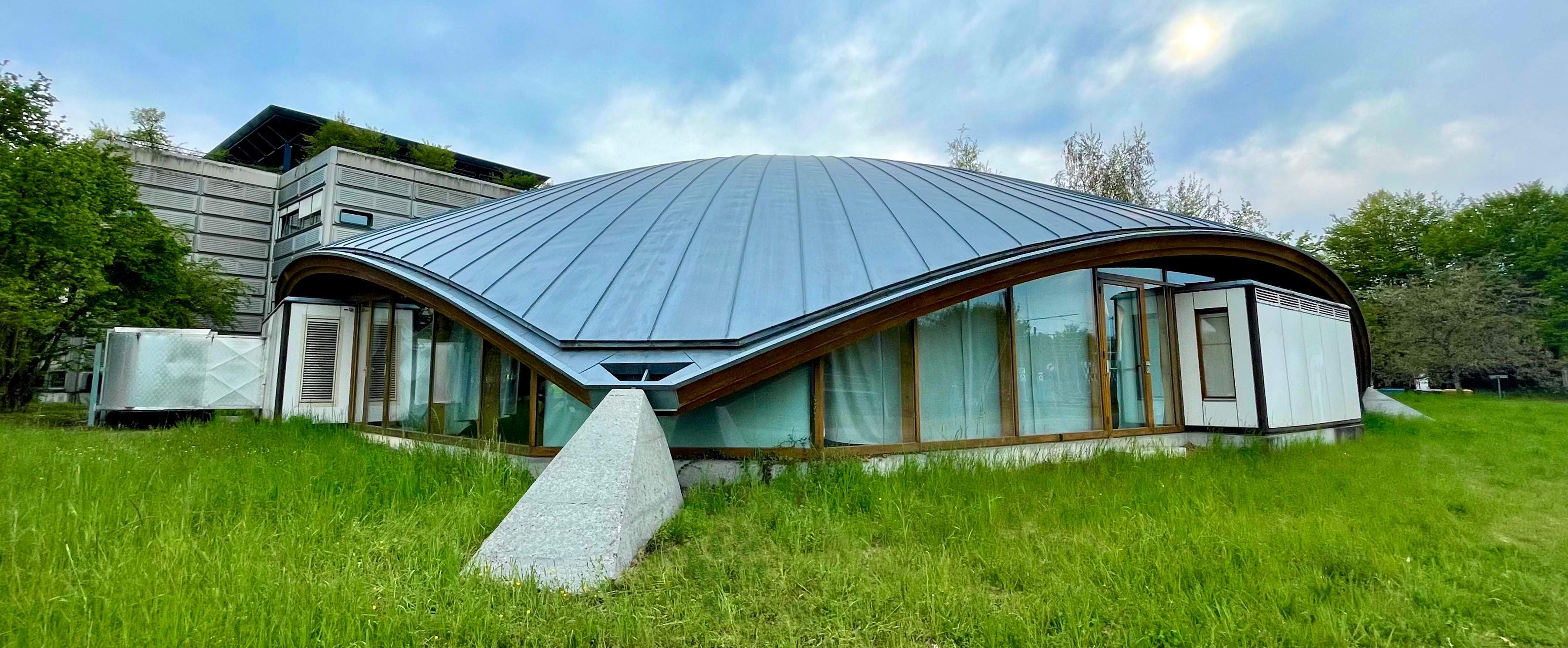}
    \end{minipage}
\end{subfigure}
\caption{The buildings where our experiments were conducted: the NEST building (left), Duebendorf, {\copyright} Zooey Braun, Stuttgart; the S\~ao Juli\~ao hospital surgery center (right, top); the Polydome lecture hall (right, bottom).}
\label{fig:nest}
\vspace{-1em}
\end{figure*}

In this work, we focus our analysis on three real-world building control experiments using distinct combinations of models and controllers: GP-based MPC~\cite{maddalena2021experimental}, bilevel DeePC~\cite{lian2021adaptive} -- relying on behavioral system theory -- and DRL matched with NN dynamical models~\cite{di2022near}. This selection is representative of the various possible choices as it covers linear, nonlinear, parametric, and non-parametric models, as well as both predictive and explicit control laws. Furthermore, these controllers were experimentally validated on different case studies, namely a hospital, a lecture hall, and an apartment, in distinct locations (see Fig.~\ref{fig:nest}). This paper hence offers a qualitative comparison of the pros and cons of each modelling and control technique through the lens of experimentalists. To the best of our knowledge, such a systematic comparison of different data-driven methods with experimental validations does not yet exist. This paper thus provides tools to help the reader decide which controller is well-suited for a particular application.

\section{Preliminaries}
\label{sec:preliminaries}

In this section, we briefly review the fundamentals of the three techniques explored in our experiments to create an intuition, and the interested reader can follow the provided pointers to key references for more detail. Throughout this work, we assume a data set of state, input, disturbance and output measurements has been gathered in $\{x_t, u_t, w_t, y_t\}_{t=1}^N$.

\subsection{Gaussian Process-based MPC}

Being grounded in Bayesian non-parametric statistics, GPs offer the user a principled way to learn an unknown function $f(x)$ from data \cite{williams2006gaussian}. 
The two principal expressions in the GP setting are the following: 
\begin{subequations}
\begin{align}
    \mu(x) &= m(x) + k_X(x)^\top (K+\sigma^2I)^{-1}(y-m_X) \label{eq.GPmean}\\
    \text{var}(x) &= k(x,x) - k_X(x)^\top (K+\sigma^2I)^{-1} k_X(x) \label{eq.GPvar}
\end{align}
\end{subequations}
where $\mu(x)$ is the learned predictive function -- the surrogate for the unknown $f(x)$ -- and $\text{var}(x)$ reflects the degree of uncertainty associated with each prediction. The choice of the kernel function $k(\cdot,\cdot)$ is central in GP modelling and gives rise to the vector $k_X(x)$ and the matrix $K$ when combined with a data set $\{x_m,y_m\}_{m=1}^M$, with $k_X(x)_i := k(x_i, x)$ and $K_{ij} := k(x_i, x_j)$ \cite{williams2006gaussian}. Lastly, $\sigma^2$ is a scalar that represents the variance of the noise affecting $y_m$.

In order to go from learning ‘static functions’ $f(x)$ to dynamical systems, one can 
employ GPs to build either autoregressive or state-space models (see e.g. \cite{maddalena2021experimental} and \cite{eleftheriadis2017identification}). We opted for the former and used the final expressions within our MPC formulation \cite{maddalena2021experimental}. 
Whereas the mean function $\mu(x)$ plays the role of the nominal dynamics, the variance information $\text{var}(x)$ is used to tighten the optimization problem constraints, conferring on it a certain degree of robustness. Although GP-MPC can be extended to incorporate online data in real-time, this feature was not exploited in our experimental investigation since it would bring about both theoretical and computational challenges \cite{lederer2020training,capone2020localized}.

\subsection{Robust Bilevel DeePC}

Behavioral system theory relies on Hankel matrices, defined as follows for a time series $s:=\{s_i\}_{i=1}^T$:
\begin{align*}
    \Han_L(s):=
    \begin{bmatrix}
    s_1 & s_2&\dots&s_{T-L+1}\\
    s_2 & s_3&\dots&s_{T-L+2}\\
    \vdots &\vdots&&\vdots\\
    s_{L} & s_{L+1}&\dots&s_T
    \end{bmatrix}\,,
\end{align*}
where $L$ is the depth of the matrix. The key idea behind any DeePC-related schemes is to use Willems' Fundamental Lemma~\cite{willems2005note} to generate trajectory predictions from a persistently excited (PE) data set (i.e. $\Han_L(u_\textbf{d})$ is full row rank, where boldface subscripts $\textbf{d}$ denotes past data from the training set). 
Under the PE condition, this lemma states that any $L$-step Input/Output (I/O) trajectory is in the column space of the stacked Hankel matrix $\Han_L(u_\textbf{d}, y_\textbf{d}):=\begin{bmatrix} \Han_{L}(u_{\textbf{d}})^\top&\Han_L(y_{\textbf{d}})^\top \end{bmatrix}^\top$.
One can hence use past data to predict future trajectories implicitly, avoiding any explicit modelling of the underlying system~\cite{coulson2019data}.
Unlike the regularized version~\cite{coulson2019data}, the trajectory prediction and the determination of control decisions were hierarchically coupled through a bilevel optimization problem in our experiments~\cite{lian2021adaptive}:
\begin{subequations}\label{eqn:rb_deepc_og}
\begin{align*}
    \min\limits_{\substack{\overline{y}_{pred}\\\overline{u}_{pred},K}} \;&\; J(y_{pred},u_{pred})
\end{align*}
\begin{align}
        \text{s.t.} \;&\forall\;w_{pred} \in \mathcal{W}\label{eqn:rb_deepc_og_wcons}\\
        &\; u_{pred}= \overline{u}_{pred}+Gw_{pred}\in \mathcal{U}\;,\label{eqn:rb_deepc_og_ucons}\\
        &\;y_{pred} = \Han_{L,pred}(y_\textbf{d})g \in \mathcal{Y}\;,\label{eqn:rb_deepc_og_ycons}\\
        &g\in\argmin_{g_l} \lVert\Han_{L,init}(y_\textbf{d})g_l-y_{init}\rVert^2+g_l^\top\mathcal{E}_g g_l\label{eqn:rb_deepc_og_lower}\\
        &\quad\quad\quad\text{s.t.}\;\begin{bmatrix}
        \Han_{L,init}(u_\textbf{d})\\\Han_{L,init}(w_\textbf{d})\\\Han_{L,pred}(u_\textbf{d})\\\Han_{L,pred}(w_\textbf{d})
        \end{bmatrix}g_l=\begin{bmatrix}u_{init}\\w_{init}\\u_{pred}\\w_{pred}
    \end{bmatrix}\nonumber\,,
\end{align}
\end{subequations}
where $u_{init},w_{init},y_{init}$ are $t_{init}$-step sequences of measured inputs, disturbances and outputs preceding the current point in time. $u_{pred}$ and $y_{pred}$ are the predicted closed-loop inputs and outputs for a specific realization of the future disturbances $w_{pred}$. 
They are therefore uncertain and have to satisfy I/O constraints $\mathcal{U}$/$\mathcal{Y}$ for any realization of $w_{pred}$ in the predictive disturbance set $\mathcal{W}$. For given realizations of $w_{pred}$ and $u_{pred}$, $y_{pred}$ is given by the lower level problem~\eqref{eqn:rb_deepc_og_ycons} and~\eqref{eqn:rb_deepc_og_lower}, where $\mathcal{E}_g$ is diagonal with decreasing value to model the preference of using recent data. 

In our experiments, the data set was robustly updated online: the input signal was randomly perturbed to guarantee the PE condition and the perturbation was treated as an auxiliary disturbance in $\mathcal{W}$. The advantages of this bilevel approach over the regularized one~\cite{coulson2019data} are detailed in~\cite{lian2021adaptive}.

\subsection{Deep Reinforcement Learning}

In Reinforcement Learning (RL), an agent interacts with the system in an iterative process: at each time step $t$, the agent chooses the control input $u_t$ based on the measurement $x_t$. The system then evolves to the next state $x_{t+1}$, and returns it to the agent, along with the step reward $r(x_t,u_t)$, and the procedure repeats until some end criterion is met. The goal of any RL algorithm is to find the optimal policy $\pi(u_t|x_t)$ that maximizes the expected returns:
\begin{align}
    J(\pi)=\mathbb{E}_\pi\left[\sum_{t=0}^\infty\gamma^tr(x_t,u_t)\right],
\end{align}
where $0<\gamma<1$ is the discount factor, balancing short- and long-term rewards. One can readily see the link with the objective function used in MPC, with the notable difference that constraints cannot be incorporated in classical RL agents and need to be learned implicitly, typically through penalties in the reward function. In practice, one usually parametrizes the policy as $\pi_\theta$ and optimizes over the parameters $\theta$. Most of the recent successes of RL were obtained through NN-based policies, which is then referred to as DRL. A plethora of algorithms to solve such (D)RL optimization problems have been developed in the past few years, and we rely on Twin Delayed Deep Deterministic policy gradients (TD3) \cite{fujimoto2018addressing} for our experiments~\cite{di2022near}.

Since classical DRL agents cannot ensure constraint satisfaction and are very data-inefficient, they cannot be deployed in a building from scratch. Indeed, they could take months or years to converge, all the while violating the comfort of the occupants \cite{wang2020reinforcement}. Consequently, in practice, we have to rely on simulation environments to train DRL agents offline before deploying them on the real system. 
In this work, we leverage Physically Consistent Neural Networks (PCNNs) to learn the temperature dynamics from data~\cite{di2021physically}. These physics-inspired NNs 
guarantee the physical consistency of the models, 
e.g. the positive correlation between heating/cooling and temperature increases/decreases, which is crucial to train meaningful DRL agents that will interact with buildings.

\section{Experimental Setups}

The common objective of the experiments described herein was to minimize the energy consumption of each building while guaranteeing thermal comfort for the occupants, i.e. maintaining indoor temperatures within predefined bounds. Note, however, that the size, occupancy profile, and actuation systems of each case study are considerably different.

\label{sec:experiments}

\subsection{Hospital Surgery Center}

The S\~ao Juli\~ao hospital surgery center, depicted in Fig.~\ref{fig:nest}, is a busy environment where over 500 surgical procedures are carried out monthly. It is located in Campo Grande, Brazil, where daily maximum temperatures reach \SI{30}{\celsius} even in winter. The building features an industrial forced-air cooling system to guarantee indoor thermal comfort. The study presented in \cite{maddalena2021experimental} was concerned with its ophthalmology section, composed of three rooms: two operating theaters and one intermediate zone. $31$ days worth of clean data were collected, sampled every \SI{10}{\minute}, and three individual GP models were then learned based roughly on 275 points each. 

The regulation task was performed over four days and the temperature constraint consisted in a single upper bound at \SI{21}{\celsius}. 
In Fig.~\ref{fig:hospital}, the temperatures and actuation profiles throughout two days are shown, and the period highlighted in grey corresponds to a chiller fault discussed in~\cite{maddalena2021experimental}. The adopted GP-MPC solution was shown to yield good performance even when subject to harsh outdoor conditions and strong internal disturbances (see \cite[Sec.~4.1]{maddalena2021experimental}). By means of simulations, it was also compared to a number of alternatives, such as ON/OFF and Proportional-Integral (PI) controllers, among others, and the results showcased the ability of GP-MPC to balance indoor thermal regulation and the chiller energy consumption. 

\input{data/Hospital}

\subsection{Polydome}\label{sect:exp_polydome}

The \textit{Polydome} is a freestanding $600 m^2$ single-zone building on the EPFL campus, regularly used as a lecture hall, that can accommodate up to 200 people (Fig.~\ref{fig:nest}). The indoor climate is maintained through a rooftop HVAC unit that we controlled for $20$ days.

Over the different experiments, we observed two main benefits of the proposed robust bilevel DeePC method over a non-robust version. Firstly, it is more robust to uncertain external weather conditions, especially when an economic objective function is considered. When the non-robust controller was deployed, we could indeed observe comfort violations around \SI{4}{\hour} and \SI{9}{\hour}, something that did not happen with the robust approach, as demonstrated in Fig.~\ref{fig:exp_polydome_3}. Furthermore, these violations led to unwanted overheating behavior later in the day around \SI{11}{\hour}.

\input{data/exp3_min_robust}

Beyond robustness, the bilevel controller also showed strong adaptivity to the working conditions. Indeed, it performed consistently along a $20$-day long experiment even though the outdoor temperature varied from \SI{15}{\celsius} to \SI{29}{\celsius}. The full plots of this experiment are available at \href{https://github.com/YingZhaoleo/Building_results}{https://github.com/YingZhaoleo/Building\_results}.

\subsection{UMAR}
\label{sec:umar}

The Urban Mining and Recycling (UMAR) unit is an apartment with two almost identical bedrooms and a living room between them in the NEST building, pictured in Fig.~\ref{fig:nest}. All the rooms are heated or cooled by letting hot or cold water run through ceiling panels, respectively, and the water is brought to the required temperature by a large reversible heat pump in the basement. During our heating experiment, the DRL agent hence controlled when to open or close the valves of the ceiling panels. 

The measurements $x$ fed to the agents regrouped autoregressive terms of room temperatures and weather conditions, as well as time information and the comfort bounds. The reward function was designed as
$r(x,u) = -\max\{T-b_u, 0\} - \max\{b_l-T,0\} - \lambda u, $
where $T$ is the indoor temperature, $b_u$ and $b_l$ represent the current upper and lower temperature bounds, and $\lambda$ balances the comfort of the occupants and the energy consumption. 
More details on the setup and simulation results can be found in \cite{di2022near}, where DRL were shown to clearly outperform industrial baselines and attain near-optimal performance.

One of the best agents was subsequently deployed in one of the bedrooms to validate its performance, and a classical hysteresis controller was deployed in the other bedroom for comparison. In Fig.~\ref{fig:exp_nest}, one can observe how the DRL agent is able to maintain the temperature close to the lower bound, thereby saving energy compared to the hysteresis controller, except when the default controller took over when the connection was lost for a few hours (shaded grey). Interestingly, despite not having access to forecasts, the agent learned to preheat the room to ensure the tightened temperature bounds are respected.

\input{data/Umar}

\section{Discussion}
\label{sec:discussion}

To begin our discussion, we want to underline that the three techniques are not in complete opposition to each other.
Indeed, one could utilize PCNNs within MPC formulations and, conversely, GPs or behavioral models could be used as simulators to train DRL agents offline. 
This also highlights the fact that, despite being often advertised as a model-free method, DRL still requires high-fidelity simulators (e.g. the models discussed herein) 
to be trained offline before being deployed on the real system to ensure adequate operations, 
a matter that is often overlooked.

The rest of this section discusses the advantages and challenges encountered using the proposed controllers on real buildings and aims at helping the reader understand which method is best suited for their application. An overview of the strengths and weaknesses of each technique is provided in Table~\ref{tab:comparison} and the details are discussed below. 

\begin{table}[]
    \centering
    \begin{tabular}{rl|c|c|c|} 
        & & \textbf{GP} & & \textbf{PCNN} \\
        & & \textbf{MPC} & \textbf{DeePC} & \textbf{DRL} \\ \hline
        \textit{A.} & Integration of nonlinearities & + & -- & + \\ 
         & Integration of prior knowledge & + & -- & + \\ \hline
        \textit{B.} & Data treatment & -- & o & + \\ 
        & Data efficiency & + & + & -- \\ \hline
        \textit{C.} & Ease of implementation  & o & -- & + \\
        & Tuning effort  & -- & + & o \\
        & Training time  & + & N/A & -- \\\hline
        \textit{D.} & Integration of constraints  & + & + & -- \\ 
        & Robustness to noise  & + & + & o \\ 
        & Robustness to unseen data  & o & + & --
        \\ \hline 
        \textit{E.} & Offline computational complexity  & -- & + & -- \\ 
        & Online computational complexity  & -- & o & + \\ \hline
        
        \textit{F.} & Interpretability  & + & + & -- \\ 
        & Need of forecasts  & -- & -- & + \\ 
         \hline
    \end{tabular}
    \caption{Qualitative comparative assessment of the different control methods.}
    \label{tab:comparison}
\end{table}

\subsection{System characteristics}

Before opting for any of the proposed methods, one should ponder over the properties of the building to be controlled and its actuation system. Whereas DeePC works well for approximately linear systems, it cannot be expected to handle strong nonlinearities as well as GP-MPC or DRL. 
In the building domain, this becomes especially apparent if the algorithms directly interact with low-level actuators, such as ventilator damper positions in forced-air systems. In practice, it is thus preferred to use DeePC for high-level control tasks, e.g. modifying temperature setpoints or
directly modulating the HVAC power consumption. 

Besides their compatibility with nonlinear dynamics, the possibility of integrating prior knowledge is another advantage of both GPs and PCNNs over DeePC. 
A concrete example would be to enforce a positive correlation between the input thermal power applied to a zone and its temperature. This is a desired property since this flaw might otherwise be exploited  
by MPC or DRL controllers, leading to spurious decisions.

\subsection{Data considerations}

In any of the three techniques, data collection and pre-processing naturally represent major workloads. Moreover, even though control performance is
heavily influenced by the data quality, it is often difficult to formally measure and quantify this influence, especially in the nonlinear setting (see e.g. \cite{lederer2020training}). The pragmatic approach is typically to gather `rich enough' data-sets around different operating points and then rely on different model validation procedures.

Since GPs have the ability to interpolate \cite{williams2006gaussian}, users have to pay particular attention to how clean the data provided to the algorithm is. On the other hand, PCNNs and NNs in general require less data cleansing and curation, but need larger batches of it to attain a good fit. Quantitatively, in our case studies, GP-MPC and DeePC were indeed much more sample efficient,
using respectively \SI{794} and \SI{192} data points, whereas 
close to $100'000$ were used to train our PCNNs, although less might suffice. 

\subsection{Training and tuning}

Once an appropriate batch of data has been gathered and cleaned, one needs to train and fine-tune the models and/or controllers. Bilvel DeePC is the least time-consuming method at this stage, especially thanks to the existence of theoretically optimal hyperparameters~\cite{lian2021adaptive}. 
Certain regularization constants used in the regularized DeePC are however still found through exhaustive search and cross-validation.
Slightly more tuning is required for GPs since the user has to experiment with different kernel functions and additional hyperparameters, such as the delays of the autoregressive structures. Finally, the 
training of PCNNs and DRL agents can 
easily be ranked as the most demanding one 
since both the model and the controller rely on NNs, which are notoriously computationally intensive to fit and involve tuning several hyperparameters. 
Additionally, since classical DRL agents do not incorporate knowledge of constraints, they have to be 
exposed to all possible scenarios during the training phase to be able to react to different operating conditions online, which might extend the training time further. 

Despite the above considerations, we want to highlight here that none of the techniques described in this paper require any structural or parametric information from the plant. In the building context, this translates to not needing access to any architectural information or construction materials, which is an important advantage over traditional white-box modelling approaches.

\subsection{Behavior during the deployment}

Some data-driven techniques are known to be brittle and might fail when faced with situations not presented to them during training.
Classical DRL agents are typically prone 
to produce spurious decisions since 
they rely on NNs, which are particularly vulnerable to unseen data~\cite{di2021physically}. 
This is aggravated by the lack of explicit constraints and might lead to temperatures diverging 
outside the desired envelope.

Conversely, DeePC and GP-MPC both take constraints into account when they compute the control actions, 
typically leading to more robust operations.
Nevertheless, if the constraints are violated for some reason (perhaps exogenous to the controller), aggressive 
responses might be triggered.
This was for example the cause of the suboptimal overheat at around \SI{10}{\hour} in the top plot of Fig.~\ref{fig:exp_polydome_3}. 
Such events might cause performance degradation, but they can often be avoided in practice, using either the uncertainty estimates provided by GPs or the robust bilevel DeePC formulation.
Note that the nature of the uncertainty considered is not the same in both cases: while GP-MPC incorporates model uncertainty, the robust bilevel DeePC characterizes external disturbances instead.

\subsection{Computational complexity}

The computational complexity 
has to be 
broken down into its offline and online components. The PCNN-DRL framework is clearly the method requiring the most computations offline, since large NNs have to be fit for both the model and the controller. Including the tuning procedure, it can easily take a few days to find good models and controllers on a GPU. By contrast, using up to \SI{300} data points, we could fit GPs in \SI{15}{\second} to \SI{2}{\minute} without GPU acceleration.
Lastly, DeePC does not entail offline computations.

Concerning the online computations, DRL agents provide control inputs almost immediately since they only require a single pass through the control NN, i.e. a few sequential matrix multiplications. On the other hand, both GP-MPC and DeePC require solving optimization problems to generate control inputs at each time step. 
DeePC boils down to Quadratic Programs (QPs), which were typically solved in under a second in our experiments, even with open-source solvers like OSQP. 
GP-MPC is surely the most computationally demanding technique at run-time as it requires solving non-convex programs with highly nonlinear equality constraints. To alleviate this load, we initialized the solver with feasible state and control trajectories (obtained with a baseline controller), and were able to compute the control actions typically within \SI{20}{\second}. 
Recall that the sampling periods in the building domain are conventionally between $5$-\SI{15}{\minute}, so that these optimizations were not an issue in practice.

\subsection{Additional comments}

Understanding the rationale behind the actions of DRL agents is not easy, especially when NN parametrizations are used. By backtracking the optimal sequences of states and controls, some explainability can be achieved in GP-MPC and DeePC, a feature that can even help users fine-tune their formulations, hence reducing the deployment time.

Remarkably, predictive control methods require forecasts of all the disturbances over the prediction horizon. Whereas weather forecasts are available on many open-source platforms, such as \href{https://www.windy.com}{windy.com}, internal heat gains are typically much harder to estimate. While good temperature predictions could be obtained without them in the Polydome, they were important to fit the GPs to the hospital data. 
On the other hand, DRL agents do not strictly require access to the aforementioned forecasts since they are able to implicitly anticipate future exogenous conditions from data. 
Nonetheless, one could easily add predictive information to the observations fed to the agents at each time step to create more robust control policies.

\section{Final Remarks}
\label{sec:conclusion}

This paper provided a pragmatic, qualitative comparison among GP-MPC, DeePC, and DRL for building control. 
In short, GPs have the great advantages of being sample efficient and yielding uncertainty estimates, but require more manual tuning than its competitors and result in more complex real-time optimization problems when paired with MPC. DeePC, on the other hand, is only well suited for linear or linearly well-approximated systems, but remains the most straighforward method to setup and deploy. Finally, the PCNN-DRL pipeline removes the need for any expert knowledge and is flexible enough to tackle many problems, but requires large amounts of data, a time-consuming offline training phase, and can cause online constraint violations.

Each approach clearly has its own merits, and they were all able to successfully handle their task. Nevertheless, arguments were put forward to show that none of the techniques can be seen as the solution to every building control problem, and an informed choice has to be made based on the plant characteristics, the amount and quality of the data, and the available 
computational power.

\bibliographystyle{IEEEtran}
\bibliography{mybib}

\end{document}

%% file: data/Hospital.tex
\definecolor{myblue}{rgb}{0.20, 0.6, 0.78}
\definecolor{mygreen}{rgb}{0.2,0.8,0.2}
\definecolor{myred}{rgb}{0.5,0,0}
\definecolor{myorange}{HTML}{e28743}
 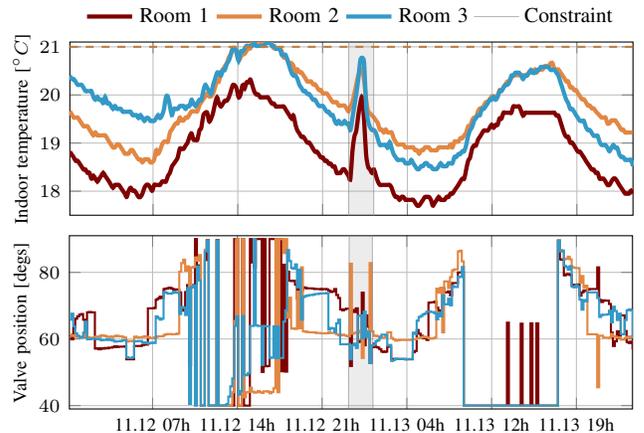
\begin{figure}
    \centering
    \begin{tikzpicture}
    \begin{axis}[
    date coordinates in=x,
    xmin=2021-11-12 00:08,
    xmax=2021-11-13 23:58,
    xtick distance=0.3,
    xticklabel=\empty,
    ymin=17.5, ymax= 21.1,
    enlargelimits=false,
    clip=true,
    grid=major,
    mark size=0.5pt,
    width=1.05\linewidth,
    height=0.45\linewidth,
    ylabel = {Indoor temperature [$^{\circ} C$]},
    ylabel style={at={(axis description cs:0.08,0.5)}},
    xlabel style={at={(axis description cs:0.5,0.05)}},
    legend columns=4,
    label style={font=\scriptsize},
    ticklabel style = {font=\scriptsize},
    legend style={
    	font=\footnotesize,
    	draw=none,
		at={(0.5,1.03)},
        anchor=south
    }    
    ]
    
    \pgfplotstableread[col sep=comma]{data/hospital.dat}{\dat}

    \addplot+ [ultra thick,smooth, mark=none, mark options={fill=white, scale=1,line width = 0.1pt},myred] table [x={t}, y={y1}] {\dat};    
    \addlegendentry{Room 1}
    \addplot+ [ultra thick,smooth, mark=none, mark options={fill=white, scale=1,line width = 0.1pt},myorange] table [x={t}, y={y2}] {\dat};    
    \addlegendentry{Room 2}
    \addplot+ [ultra thick,smooth, mark=none, mark options={fill=white, scale=1,line width = 0.1pt},myblue] table [x={t}, y={y3}] {\dat};    
    \addlegendentry{Room 3}    
    \addplot+ [name path =A, ultra thin, gray,draw opacity=0.6,mark = none] table [x={t}, y expr=\thisrow{mode}*150] {\dat};
    \addplot+ [name path =B, ultra thin, gray,draw opacity=0.6,mark = none] table [x={t}, y expr=\thisrow{mode}*-100] {\dat};
    \addplot[gray,fill opacity =0.15] fill between[of=A and B];

    \addplot+ [thick,brown,dashed,mark = none] table [x={t}, y={ymax}] {\dat};
    \addlegendentry{Constraint}     
    
    \end{axis}
    \end{tikzpicture}\\
    \begin{tikzpicture}
    \begin{axis}[
    date coordinates in=x,
    xmin=2021-11-12 00:08,
    xmax=2021-11-13 23:58,
    xtick distance=0.3,
    xticklabel=\month.\day\ \hour {h},
    ymin=39, ymax= 91,
    enlargelimits=false,
    clip=true,
    grid=major,
    mark size=0.5pt,
    width=1.05\linewidth,
    height=0.45\linewidth,
    ylabel = {Valve position [degs]},
    ylabel style={at={(axis description cs:0.08,0.5)}},
    xlabel style={at={(axis description cs:0.5,0.05)}},
    legend columns=4,
    label style={font=\scriptsize},
    ticklabel style = {font=\scriptsize},
    legend style={
    	font=\footnotesize,
    	draw=none,
		at={(0.5,1.03)},
        anchor=south
    }    
    ]
    
    \pgfplotstableread[col sep=comma]{data/hospital.dat}{\dat}

    \addplot+ [thick,const plot, mark=none, mark options={fill=white, scale=1,line width = 0.1pt},myred] table [x={t}, y={u1}] {\dat};    
    \addplot+ [thick,const plot, mark=none, mark options={fill=white, scale=1,line width = 0.1pt},myorange] table [x={t}, y={u2}] {\dat};    
    \addplot+ [thick,const plot, mark=none, mark options={fill=white, scale=1,line width = 0.1pt},myblue] table [x={t}, y={u3}] {\dat};    
    \addplot+ [name path =A, ultra thin, gray,draw opacity=0.6,mark = none] table [x={t}, y expr=\thisrow{mode}*150] {\dat};
    \addplot+ [name path =B, ultra thin, gray,draw opacity=0.6,mark = none] table [x={t}, y expr=\thisrow{mode}*-100] {\dat};
    \addplot[gray,fill opacity =0.15] fill between[of=A and B];

    \end{axis}
    \end{tikzpicture}
   
    \caption{Hospital November $11^{th}-12^{th}$, $2021$: three rooms controlled by the GP-MPC controller.}
    \label{fig:hospital}        
 \end{figure}

%% file: data/exp3_min_robust.tex
\definecolor{myblue}{rgb}{0.20, 0.6, 0.78}
\definecolor{mygreen}{rgb}{0.2,0.8,0.2}
\definecolor{myred}{rgb}{0.5,0,0}
 \begin{figure}
    \centering
        \begin{tikzpicture}
    \begin{axis}[
    date coordinates in=x,
    xmin=2021-05-13 23:59,
    xmax=2021-05-14 24:00,
    xtick distance=0.125,
    xticklabel=\empty,
    ymin=5, ymax= 30.5,
    enlargelimits=false,
    clip=true,
    grid=major,
    mark size=0.5pt,
    width=1\linewidth,
    height=0.45\linewidth,
    ylabel = {Indoor temperature [$^{\circ} C$]},
    ylabel style={at={(axis description cs:0.085,0.5)}},
    xlabel style={at={(axis description cs:0.5,0.05)}},
    legend columns=3,
    label style={font=\scriptsize},
    ticklabel style = {font=\scriptsize},
    legend style={
    	font=\footnotesize,
    	draw=none,
		at={(0.5,1.03)},
        anchor=south
    }    
    ]
    
    \pgfplotstableread[col sep=comma]{data/polydome_min.dat}{\dat}

    \addplot+ [ultra thick,smooth, mark=none, mark options={fill=white, scale=1,line width = 0.1pt},myred] table [x={tt}, y={y}] {\dat};    
    \addlegendentry{System output: y}
    
    \addlegendimage{line legend,ultra thick,smooth, mark=none, mark options={fill=white, scale=1,line width = 0.2pt}, myblue}
    \addlegendentry{Control input: u}       
    
    \addplot+ [thick,brown,dashed,mark = none] table [x={tt}, y={max}] {\dat};
    \addplot+ [thick,brown, dashed, mark = none] table [x={tt}, y={min}] {\dat};
    \addlegendentry{Constraint}     
    
    \end{axis}
    
    \begin{axis}[
    axis y line*=right,
    ymin=-0.125, ymax= 3.2,
    ylabel = {Electrical power [$kWh$]},
    axis x line=none,
    date coordinates in=x,
    xmin=2021-05-13 23:59,
    xmax=2021-05-14 24:00,
    xtick distance=0.125,
    xticklabel=\hour:\minute,
    enlargelimits=false,
    mark size=0.5pt,
    width=1\linewidth,
    height=0.45\linewidth,
    legend style={
    	font=\footnotesize,
    	draw=none,
		at={(0.5,1.00)},
        anchor=south
    },
    ylabel style={at={(axis description cs:1.25,0.5)}},
    legend columns=2,
    label style={font=\scriptsize},
    ticklabel style = {font=\scriptsize}
    ]
    
    \pgfplotstableread[col sep=comma]{data/polydome_min.dat}{\dat}
   
    \addplot+ [ultra thick, const plot, mark=none, mark options={fill=white, scale=1.5,line width = 0.2pt}, myblue] table [x={tt}, y expr=\thisrow{u}*0.25] {\dat};
      
    \end{axis} 
    
    \end{tikzpicture}\\
    \begin{tikzpicture}
    \begin{axis}[
    date coordinates in=x,
    xmin=2021-05-24 23:59,
    xmax=2021-05-25 24:00,
    xtick distance=0.125,
    xticklabel=\hour:\minute,
    ymin=5, ymax= 30.5,
    enlargelimits=false,
    clip=true,
    grid=major,
    mark size=0.5pt,
    width=1\linewidth,
    height=0.45\linewidth,
    ylabel = {Indoor temperature [$^{\circ} C$]},
    xlabel= Time,
    ylabel style={at={(axis description cs:0.085,0.5)}},
    xlabel style={at={(axis description cs:0.5,0.05)}},
    legend columns=3,
    label style={font=\scriptsize},
    ticklabel style = {font=\scriptsize},
    legend style={
    	font=\footnotesize,
    	draw=none,
		at={(0.5,1.03)},
        anchor=south
    }    
    ]
    
    \pgfplotstableread[col sep=comma]{data/polydome_min_robust.dat}{\dat}

    \addplot+ [ultra thick,const plot, mark=none, mark options={fill=white, scale=1,line width = 0.1pt},myred] table [x={tt}, y={y}] {\dat};    
    
    \addlegendimage{line legend,ultra thick,smooth, mark=none, mark options={fill=white, scale=1,line width = 0.2pt}, myblue}
    
    \addplot+ [thick,brown,dashed,mark = none] table [x={tt}, y={max}] {\dat};
    \addplot+ [thick,brown, dashed, mark = none] table [x={tt}, y={min}] {\dat};
    
    \end{axis}
    
    \begin{axis}[
    axis y line*=right,
    ymin=-0.125, ymax= 3.2,
    ylabel = {Electrical power [$kWh$]},
    axis x line=none,
    date coordinates in=x,
    xmin=2021-05-24 23:59,
    xmax=2021-05-25 24:00,
    xtick distance=0.125,
    xticklabel=\hour:\minute,
    enlargelimits=false,
    mark size=0.5pt,
    width=1\linewidth,
    height=0.45\linewidth,
    legend style={
    	font=\footnotesize,
    	draw=none,
		at={(0.5,1.00)},
        anchor=south
    },
    ylabel style={at={(axis description cs:1.25,0.5)}},
    legend columns=2,
    label style={font=\scriptsize},
    ticklabel style = {font=\scriptsize}
    ]
    
    \pgfplotstableread[col sep=comma]{data/polydome_min_robust.dat}{\dat}
   
    \addplot+ [ultra thick,const plot, mark=none, mark options={fill=white, scale=1.5,line width = 0.2pt}, myblue] table [x={tt}, y  expr=\thisrow{u}*0.25] {\dat};
      
    \end{axis} 
    
    \end{tikzpicture}\\
    \caption{\textit{Polydome}. Top: May $14^{th}$, 2021: non-robust bilevel data-driven control. Bottom: May $25^{th}$, 2021: robust bilevel  data-driven  control}
    \label{fig:exp_polydome_3}        
 \end{figure}
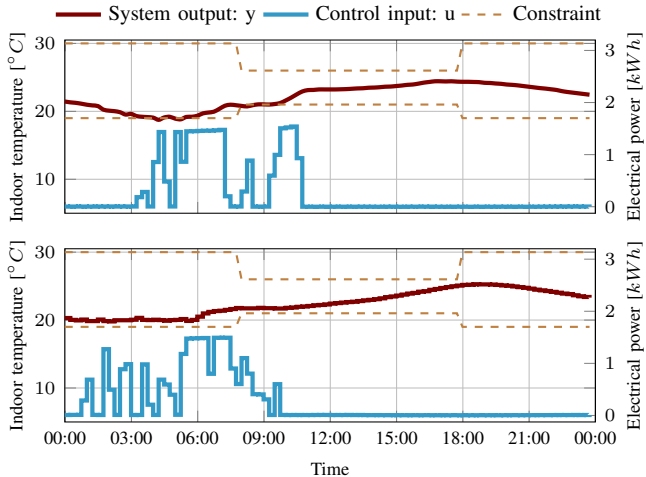

%% file: data/Umar.tex
 \definecolor{myblue}{rgb}{0.20, 0.6, 0.78}
\definecolor{mygreen}{rgb}{0.2,0.8,0.2}
\definecolor{myred}{rgb}{0.5,0,0}
 \begin{figure}
    \centering
    \begin{tikzpicture}
    \begin{axis}[
    date coordinates in=x,
    xmin=2022-02-15 12:00,
    xmax=2022-02-18 08:00,
    xtick distance=0.5,
    xticklabel=\empty,
    ymin=20.9, ymax= 24.2,
    enlargelimits=false,
    clip=true,
    grid=major,
    mark size=0.5pt,
    width=1.05\linewidth,
    height=0.45\linewidth,
    ylabel = {Indoor temperature [$^{\circ} C$]},
    ylabel style={at={(axis description cs:0.085,0.5)}},
    xlabel style={at={(axis description cs:0.5,0.05)}},
    legend columns=3,
    label style={font=\scriptsize},
    ticklabel style = {font=\scriptsize},
    legend style={
    	font=\footnotesize,
    	draw=none,
		at={(0.5,1.03)},
        anchor=south
    } 
    ]
    
    \pgfplotstableread[col sep=comma]{data/UMAR.dat}{\dat};

    \addplot+ [ultra thick,smooth, mark=none, mark options={fill=white, scale=1,line width = 0.1pt},myred] table [x={t}, y={ybar}] {\dat};    
    \addlegendentry{Bang-bang Control};
    \addplot+ [ultra thick,smooth, mark=none, mark options={fill=white, scale=1,line width = 0.1pt},myblue] table [x={t}, y={y}] {\dat};    
    \addlegendentry{DRL Agent};
    \addplot+ [thick,brown,dashed,mark = none,forget plot] table [x={t}, y={ymax}] {\dat};
    \addplot+ [thick,brown,dashed,mark = none] table [x={t}, y={ymin}] {\dat};
    \addlegendentry{Contstraints}
    
    \addplot+ [name path =A, ultra thin, gray,draw opacity=0.6,mark = none] table [x={t}, y expr=\thisrow{mode}*150] {\dat};
    \addplot+ [name path =B, ultra thin, gray,draw opacity=0.6,mark = none] table [x={t}, y expr=\thisrow{mode}*-100] {\dat};
    \addplot[gray,fill opacity =0.15] fill between[of=A and B];

    \end{axis}
    \end{tikzpicture}   
    \begin{tikzpicture}
    \begin{axis}[
    date coordinates in=x,
    xmin=2022-02-15 12:00,
    xmax=2022-02-18 08:00,
    xtick distance=0.5,
    xticklabel=\month.\day\ \hour {h},
    ymin=-1, ymax= 103,
    enlargelimits=false,
    clip=true,
    grid=major,
    mark size=0.5pt,
    width=1.05\linewidth,
    height=0.45\linewidth,
    ylabel = {Valve position [$\%$]},
    xlabel= Time,
    ylabel style={at={(axis description cs:0.088,0.5)}},
    xlabel style={at={(axis description cs:0.5,0.05)}},
    legend columns=3,
    label style={font=\scriptsize},
    ticklabel style = {font=\scriptsize}   
    ]
    
    \pgfplotstableread[col sep=comma]{data/UMAR.dat}{\dat};

    \addplot+ [ultra thick,const plot, mark=none, mark options={fill=white, scale=1,line width = 0.1pt},myred] table [x={t}, y expr=\thisrow{ubar}*100] {\dat};    
    \addplot+ [ultra thick,const plot, mark=none, mark options={fill=white, scale=1,line width = 0.1pt},myblue] table [x={t}, y expr=\thisrow{u}*100] {\dat};
    \addplot+ [name path =A, ultra thin, gray,draw opacity=0.6,mark = none] table [x={t}, y expr=\thisrow{mode}*150] {\dat};
    \addplot+ [name path =B, ultra thin, gray,draw opacity=0.6,mark = none] table [x={t}, y expr=\thisrow{mode}*-100] {\dat};
    \addplot[gray,fill opacity =0.15] fill between[of=A and B];

    \end{axis}
    \end{tikzpicture} 
    \caption{UMAR, February $15^{th}-18^{th}$, 2022: one bedroom controlled by a classical rule-based controller (red), the other by the proposed DRL agent (blue). The connection was interrupted in the shaded area and default controllers took over.}
\label{fig:exp_nest}
 \end{figure}
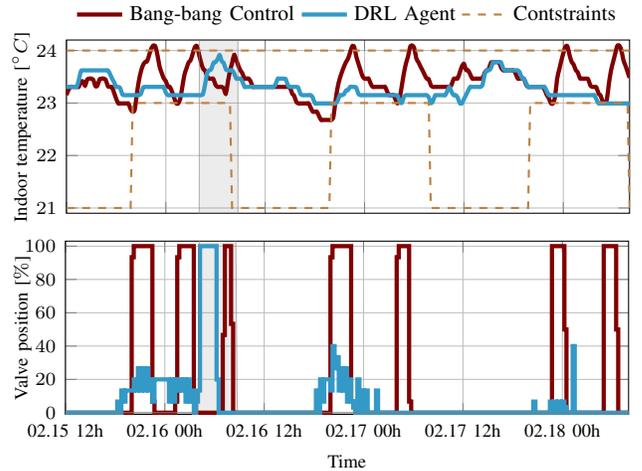